\documentclass[english]{article}
\usepackage[T1]{fontenc}
\usepackage[latin1]{inputenc}
\usepackage{a4wide}
\usepackage{float}
\usepackage{amsmath}
\usepackage{graphicx}
\usepackage{setspace}
\usepackage{amssymb}

\makeatletter

\providecommand{\tabularnewline}{\\}

\usepackage{babel}
\makeatother
\begin{document}
\vspace*{2.cm}\LARGE\thispagestyle{empty}

\begin{center}\textbf{\LARGE Statistical Pattern Recognition: \\
Application to $\nu_{\mu}\rightarrow\nu_{\tau}$ Oscillation Searches \\ 
Based on Kinematic 
Criteria}\end{center}{\LARGE \par}

\normalsize\vspace*{0.3cm}

\begin{center}A. Bueno%
\footnote{a.bueno@ugr.es%
}, A. Martínez de la Ossa%
\footnote{ossa@ugr.es%
} and S. Navas-Concha %
\footnote{navas@ugr.es%
}

Departamento de Física Teórica y del Cosmos \& C.A.F.P.E.,
University of Granada, Spain\end{center}

\begin{center}A. Rubbia%
\footnote{andre.rubbia@cern.ch%
}

Institute for Particle Physics, ETH H\"onggerberg, Z\"urich, Switzerland

\end{center}

\vspace*{1.5cm}

\begin{abstract}
Classic statistical techniques (like the multi-dimensional likelihood
and the Fisher discriminant method) together with Multi-layer Perceptron and 
Learning Vector Quantization 
Neural Networks have been systematically used in order to find the best sensitivity
when searching for $\nu_\mu \to \nu_{\tau}$ oscillations. 
We discovered that for a general direct $\nu_\tau$ appearance search based 
on kinematic criteria:  
a) An optimal discrimination power 
is obtained using only three variables ($E_{visible}$, $P_{T}^{miss}$ and $\rho_{l}$) 
and their correlations. Increasing the number of variables (or combinations of variables)
only increases the complexity of the problem, but does not result
in a sensible change of the expected sensitivity. 
b) The multi-layer perceptron approach offers the best performance. 
As an example to assert numerically those points, 
we have considered the problem of $\nu_\tau$ appearance at the CNGS 
beam using a Liquid Argon TPC detector.


\newpage\pagestyle{plain} \setcounter{page}{1} \setcounter{footnote}{0}
\end{abstract}

\section{Introduction}
\label{sec:one} 

The experimental confirmation that atmospheric and solar neutrinos do 
oscillate \cite{superk, sno}, and therefore have mass, represents the first 
solid clue for the existence of new physics beyond the Standard Model \cite{SM}. 
Results from experiments carried out with neutrinos 
produced in artificial sources, like reactors and accelerators, strongly support 
the fact that neutrinos are massive \cite{K2K,Kamland}. 

Notwithstanding the impressive results achieved by current experiments, neutrino phenomenology 
is a very rich and active field, where plenty of open questions still await for a definitive answer. 
Thus, many next-generation neutrino experiments are being designed and proposed to 
measure with precision the parameters that govern the oscillation (mass differences 
and mixing angles) \cite{wark}. New facilities like super-beams, beta beams \cite{zucchelli} 
and neutrino factories \cite{geer} have been put forward and their performances studied in detail 
in order to ascertain whether they can give an answer to two fundamental questions: what is 
the value of the mixing angle between the first and the third family, 
and whether CP violation takes place in the leptonic sector \cite{nf}. 

Recently, the Super-Kamiokande Collaboration 
has measured a first evidence of the sinusoidal behaviour of neutrino disappearance 
as dictated by neutrino oscillations \cite{lovere}. However, although the most favoured hypothesis 
for the observed $\nu_\mu$ disappearance 
is that of $\nu_\mu \to \nu_\tau$ oscillations, no direct evidence for $\nu_\tau$ appearance 
exists up to date. A long baseline neutrino beam, 
optimized for the parameters favoured by atmospheric oscillations, 
has been approved in Europe to look for explicit $\nu_{\tau}$ appearance: 
the CERN-Laboratori Nazionali del Gran Sasso (CNGS) beam \cite{BEAM}. 
The approved experimental program consists of two experiments ICARUS
\cite{ICARUS} and OPERA \cite{OPERA} that will search for 
$\nu_\mu \to \nu_\tau$ oscillations using complementary techniques.


Given the previous experimental efforts \cite{NOMAD,CHORUS} and present interest
in direct $\nu_{\tau}$ appearance, we assess in this note the
performance of several statistical techniques
applied to the search for $\nu_{\tau}$ using
kinematic techniques. Classic statistical methods (like multi-dimensional likelihood and Fischer's
discriminant schemes) and \emph{Neural Networks} based ones (like
multi-layer perceptron and self-organized neural networks) have been
applied in order to find the approach that offers the best sensitivity. 

\section{Oscillation Search Using Kinematic Criteria}

The original proposal to observe for the first time the direct
appearance of a $\nu_\tau$ by means of kinematic criteria dates back to 1978 \cite{Albright:1978ni}. 
Based on the capabilities to measure the direction of the hadronic
jet, the interaction of the neutrino associated with the tau lepton
can be spotted thanks to: a) the presence of a sizable missing transverse
momentum; b) certain angular correlations between the direction of the
prompt lepton and the hadronic jet, in the plane transverse to the
incoming neutrino beam direction.

NOMAD \cite{NOMAD} was a pioneering experiment in the use of kinematic
criteria applied to a $\nu_\mu\to\nu_\tau$ oscillation search. 
The kinematic approach was validated after several years of successful 
operation at the CERN WANF neutrino beam \cite{nomadosc,
Astier:1999vc}. This short-baseline experiment set the most competitive limit for 
$\nu_\mu\to\nu_\tau$ oscillations at high values of $\Delta m^2$
\cite{Astier:2001}.

An impressive background rejection power {\cal
O}($10^5$) was needed in NOMAD. To achieve this, a multidimensional likelihood was built
taking advantage of: on the one hand, the different event kinematics for signal and
background events; on the other, the existing correlations among
the variables used. To further enhance the sensitivity, the signal
region was divided into several bins. 

Given the interest that $\nu_\mu\to\nu_\tau$ oscillation searches have
nowadays for the region of $\Delta m^2 \sim 10^{-3}$ eV$^2$, we have
considered the problem of finding the statistical approach that offers
the best sensitivity for this kind of search. Unlike NOMAD, we do not
try to improve the sensitivity by splitting the signal regions into a
set of independent bins.

We have simply compared the discrimination power offered by a multi-dimensional
likelihood, the Fisher discriminant method and a neural network. 
As a general conclusion, we have observed that neural networks offer the
best background rejection power thanks to their ability to find complex correlations
among the kinematic variables. In addition, they allow to reduce the
complexity of the problem, given that a small number of input
variables is enough to optimize the experimental sensitivity. 
These conclusions are valid for direct $\nu_\tau$ appearance searches
performed either with atmospheric or accelerator neutrinos. 
In what follows we give a numerical example that illustrates the conclusions of this study.

\section{Detector Configuration and Data Simulation}
\label{sec:simulation}

To obtain a numerical evaluation of the performances of the different
statistical techniques we used, and assess which of them gives the
best sensitivity when searching for direct $\nu_{\tau}$
appearance by means of kinematic criteria, we have considered the particular case of the CNGS
beam. 

We assume a detector configuration consisting of 3 ktons of Liquid Argon \cite{ICARUS}. 
In our simulation the total mass of active (imaging) Argon amounts to 2.35 ktons. 
We assumed five years running of the CNGS beam
in shared mode ($4.5\times10^{19}$ p.o.t. per year), which translates
into a total exposure of $5\times 2.35=11.75$ kton$\times$year.
The total event rates expected are 252 (17) $\nu_e \ (\bar\nu_e)$ CC
events and 50 $\nu_\tau$ CC events with the $\tau$ decaying into an
electron plus two neutrinos (we assume maximal mixing and 
$\Delta m_{23}^{2}=3\times10^{-3}$ eV$^{2}$; these values are
compatible with the allowed range given by atmospheric neutrinos).
Before cuts, the signal over background ratio, in active LAr, is $50/252\simeq0.2$.

The study of the capabilities to reconstruct and analyze high-energy neutrino
events was done using fully simulated $\nu_{e}$CC events inside the
whole LAr active volume. Neutrino cross
sections and the generation of neutrino interactions is based on the
NUX code \cite{NUX}; final state particles are then tracked using
the FLUKA package \cite{FLUKA}. The angular and energy resolutions 
used in the simulation of final state electrons and individual hadrons are 
identical to those quoted in \cite{ICARUS}.

In order to apply the most efficient kinematic selection, it is mandatory
to reconstruct with the best possible resolution the energy and the
angle of the hadronic jet and the prompt lepton, with particular attention
to the tails of the distributions. Therefore, the energy flow algorithm 
has been designed with care, taking into account the
needs of the tau search analysis. 

The ability to look for tau appearance events is limited by the containment
of high energy neutrino events. Energy leakage outside the active
imaging volume creates tails in the kinematic variables that fake
the presence of neutrinos in the final state. We therefore impose
fiducial cuts in order to guarantee that on average the events will
be sufficiently contained.

The fiducial volume is defined by looking at the profiles of the total
missing transverse momentum and of the total visible energy of the
events. The average value of these variables is a good estimator of
how much energy is leaking on average. After fiducial cuts, we keep  
65\% of the total number of events occurring in the active LAr volume. 
This means a total exposure of 7.6 kton$\times$year after five years 
of shared CNGS running.

Table \ref{tab:mcevents} summarizes the total amount of simulated
data used for this study. 
We note that $\nu_{\tau} \ (\nu_{e})$ CC sample, generated
in active LAr, is more than a factor 250 (50) larger than the expected
number of collected events after five years of CNGS running. 

%
%
%
%

%
\begin{table}[H]
\begin{center}\begin{tabular}{|c|c|c|}
\hline 
\textbf{Process}&
\textbf{$\nu_{e}$CC}&
\textbf{$\nu_{\tau}$CC }\tabularnewline
&
&
\textbf{($\tau\rightarrow e$)}\tabularnewline
\hline
\hline 
\textbf{Active LAr}&
14200 [252]&
13900 [50]\tabularnewline
\hline 
\textbf{Fiducial Vol.}&
\textbf{\emph{9250}}[163]&
\textbf{\emph{9000}}[33]\tabularnewline
\hline
\end{tabular}\end{center}

\caption{Amount of fully generated data in Active and Fiducial LAr volumes. 
Between brackets we show the
expected number of events after five years of data taking at CNGS with
a 3 kton detector.}

\label{tab:mcevents}
\end{table}

\section{Statistical Pattern Recognition Applied to Oscillation Searches\label{sec:Statistical-Pattern-Recognition}}

In the case of a $\nu_{\mu}\rightarrow\nu_{\tau}$ oscillation search with Liquid Argon, 
the golden channel to look for $\nu_{\tau}$ appearance is the decay
of the tau into an electron and a pair neutrino anti-neutrino due
to: (a) the excellent electron identification capabilities; (b) the
low background level, since the intrinsic $\nu_{e}$ and $\bar{\nu}_{e}$
charged current contamination of the beam is at the level of one per
cent.

Kinematic identification of the $\tau$ decay \cite{Astier:1999vc}, which follows the
$\nu_{\tau}$CC interaction, requires excellent detector performance:
good calorimetric features together with tracking and topology reconstruction
capabilities. In order to separate $\nu_{\tau}$ events from the background,
a basic criteria can be used: an unbalanced total transverse momentum due to neutrinos produced
in the $\tau$ decay.

\begin{figure}
\begin{center}\includegraphics[%
  width=10cm,
  keepaspectratio]{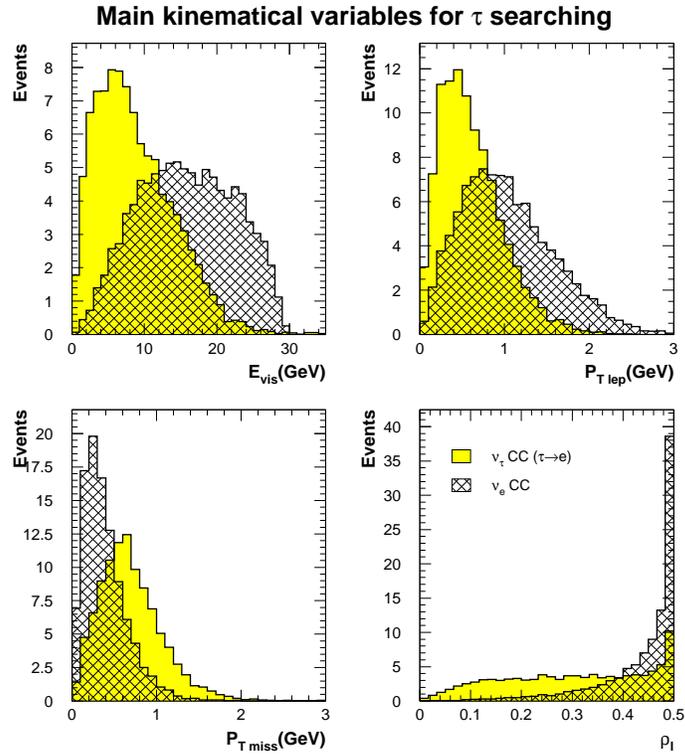}\end{center}

\caption{Visible energy (top left), transverse missed momentum (bottom left),
transverse electron momentum (top right) and $\rho_{l}$ (bottom right).
Histograms have an arbitrary normalization.}

\label{fig:mainvars}
\end{figure}

In figure \ref{fig:mainvars} we illustrate the difference on kinematics
for signal and background events. We plot four of the most discriminating
variables: 
\begin{itemize} 
\item $E_{vis}$: Visible energy. 
\item $P_{T}^{miss}$: Missing momentum in the transverse plane with respect to the direction of the 
incident neutrino beam. 
\item $P_{T}^{lep}$: Transverse momentum of the prompt electron candidate. 
\item $\rho_{l}=\frac{{\textstyle P_{T}^{lep}}}{{\textstyle P_{T}^{lep}+P_{T}^{had}+P_{T}^{miss}}}$%
\end{itemize}
 Signal events tend to accumulate in low $E_{vis}$, low
$P_{T}^{lep}$, low $\rho_{l}$ and high $P_{T}^{miss}$ regions.

Throughout this
article, we take into account only the background due to electron neutrino 
charged current interactions. Due to the low content the beam has on $\bar{\nu}_e$, 
charged currents interactions 
of this type have been observed to give a negligible contribution to the total expected background. 
We are
confident that neutral current background can be reduced to a negligible level
using LAr imaging capabilities and algorithms based on the different
energy deposition showed by electrons and $\pi^{0}$(see for example~\cite{pi0}). 
Therefore it will not be further considered. The contamination due to charm production 
and $\nu_\mu$ CC events, where the prompt muon is not identified as such, was studied 
by the ICARUS Collaboration \cite{ICANOE} and showed to be less important 
than $\nu_{e}$CC background.

\subsection{Oscillation Search Using Classic Statistical Methods}

\subsubsection{The Multi-dimensional Likelihood\label{sub:likelihood}}

The first method adopted for the $\tau$ appearance search is
the construction of a multi-dimensional likelihood function (see for
example \cite{Cowan}), which is used as the unique discriminant between
signal and background. This approach is, a priori, an optimal discrimination tool since 
it takes into account correlations between the chosen variables.

A complete likelihood function should contain five variables (three
providing information of the plane normal to the incident neutrino
direction and two more providing longitudinal information). However,
in a first approximation, we limit ourselves to the discrimination information
provided by the three following variables: $E_{visible}$, $P_{T}^{miss}$ and $\rho_l$.

As we will see later, all the discrimination power is contained in
these variables, therefore we can largely reduce the complexity of
the problem without affecting the sensitivity of the search. Two likelihood
functions were built, one for $\tau$ signal ($\mathcal{L_{S}}$)
and another for background events ($\mathcal{L_{B}}$). The discrimination
was obtained by taking the ratio of the two likelihoods: 

\begin{equation}
ln(\lambda)\equiv\mathcal{L}([E_{visible},P_{T}^{miss},\rho_{l}])=\frac{\mathcal{L_{S}}([E_{visible},P_{T}^{miss},\rho_{l}])}{\mathcal{L_{B}}([E_{visible},P_{T}^{miss},\rho_{l}])}\label{eq:likelihood}\end{equation}

In order to avoid a bias in our estimation, half of the generated data was used to build 
the likelihood functions and the other 
half was used to evaluate overall efficiencies. 
Full details about the multi-dimensional likelihood algorithm can
be found elsewhere \cite{pi0}. However, we want to point out here some
important features of the method.

A partition of the hyperspace of input variables is required: The
multi-dimensional likelihood will be, in principle, defined over a
lattice of bins. The number of bins to be filled when constructing
likelihood tables grows like $n^{d}$ where $n$ is the number of
bins per variable and $d$ the number of these variables. This leads
to a ``dimensionality'' problem when we increment the number of
variables, since the amount of data required to have a well defined
value for ln$\lambda$ in each bin of the lattice will grow exponentially. 

\begin{figure}
\begin{center}\includegraphics[%
  width=10cm,
  keepaspectratio]{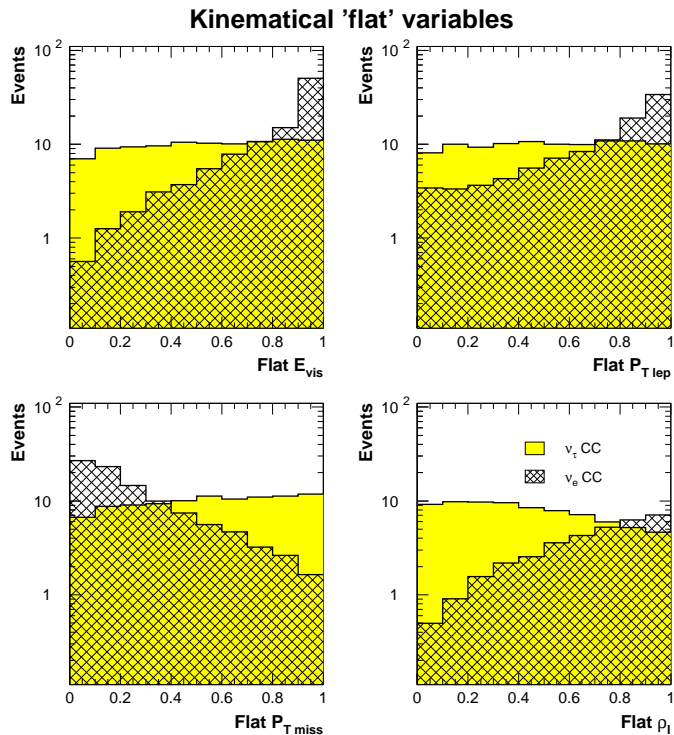}\end{center}

\caption{Comparison for {}``flat'' $E_{visible}$, $P_{T}^{lep}$, $\rho_{l}$
and $P_{T}^{miss}$ variables between $\tau$ signal and $\nu_{e}$
CC events. Arbitrary normalization has been taken into account when
plotting background events.\label{fig:flat}}
\end{figure}

In order to avoid regions populated with very few events, input variables
must be redefined to have the signal uniformly distributed in the
whole input hyperspace, hence $E_{visible}$, $P_{T}^{miss}$ and
$\rho_{l}$ are replaced by ``flat'' variables (see figure \ref{fig:flat}).
Besides, an adequate smoothing algorithm is needed in order to alleviate
fluctuations in the distributions in the hyperspace and also, to provide
a continuous map from the input variables to the multi-dimensional
likelihood one (ln$\lambda$).

\begin{figure}
\begin{center}\includegraphics[%
  width=12cm,
  keepaspectratio]{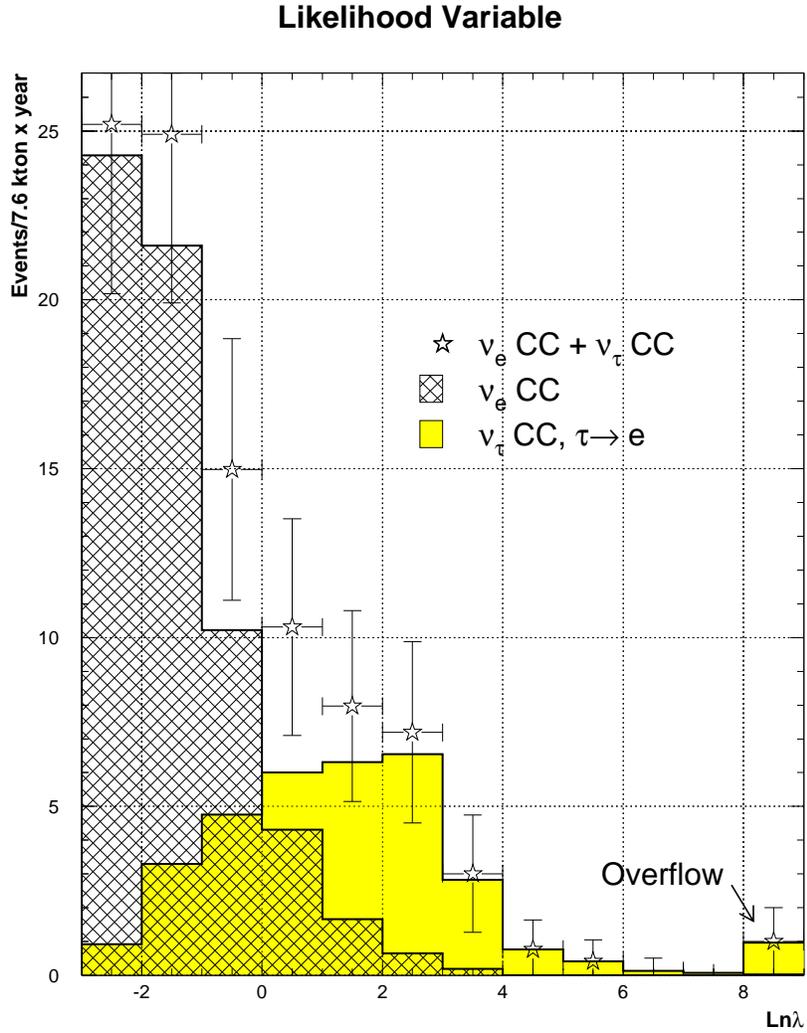}\end{center}

\caption{Multi-dimensional likelihood distributions for $\nu_{e}$ CC and
$\tau\rightarrow e$ events. The last bin in signal includes the event
overflow. Error bars in $\nu_{e}$CC + $\nu_{\tau}$CC sample represent
statistical fluctuations in the expected profile measurements after
5 years of data taking with shared running CNGS and a 3 kton detector
configuration.\label{fig:likes}}
\end{figure}

Ten bins per variable were used, giving rise to a total of $10^{3}$
bins. Figure \ref{fig:likes} shows the likelihood distributions for
background and tau events assuming five years running of CNGS (total
exposure of 7.6 kton $\times$ year for events occurring inside the
fiducial volume). %

Table \ref{tab:lkl} shows, for different cuts of $\ln\lambda$, the
expected number of tau and $\nu_{e}$ CC background events. As reference
for future comparisons, we focus our attention in the cut ln$\lambda>1.8$.
It gives a signal selection efficiency around 25\% (normalized to
the total number of $\tau$ events in active LAr). This $\tau$ efficiency
corresponds to 12.9 signal events. For this cut, we expect $1.1\pm0.2$
background events. After cuts are imposed, this approach predicts
a S/B ratio similar to 13.
\begin{table}
\begin{center}\begin{tabular}{|l|c|c|c|}
\hline 
 &
$\nu_{\tau}$CC $(\tau\rightarrow e)$ &
&
 $\nu_{\tau}$ CC $(\tau\rightarrow e)$ \tabularnewline
~~~~~~~~Cuts &
Efficiency &
$\nu_{e}$CC&
 $\Delta m^{2}=$\tabularnewline
&
 ($\%$) &
&
 $3\times10^{-3}$ eV$^{2}$\tabularnewline
\hline
Initial &
 100 &
 252 &
 50 \tabularnewline
\hline
Fiducial volume &
 65 &
 163 &
 33 \tabularnewline
\hline
$\ln\lambda>0.0$&
 48 &
 $6.8\pm0.5$ &
 $24.0\pm0.6$\tabularnewline
 $\ln\lambda>0.5$&
 42 &
$3.6\pm0.3$&
 $20.8\pm0.6$\tabularnewline
 $\ln\lambda>1.0$&
 36 &
$2.5\pm0.3$&
$18.0\pm0.6$\tabularnewline
 $\ln\lambda>1.5$&
 30 &
 $1.7\pm0.2$ &
$15.2\pm0.5$\tabularnewline
\textbf{ln $\mathbf{\lambda>1.8}$}&
\textbf{25}&
\textbf{$\mathbf{1.1\pm0.2}$}&
\textbf{$\mathbf{12.9\pm0.5}$}\tabularnewline
 ln $\lambda>2.0$&
 23&
$0.86\pm0.16$&
$11.7\pm0.5$\tabularnewline
 $\ln\lambda>2.5$&
 16 &
 $0.40\pm0.12$ &
 $8.1\pm0.4$ \tabularnewline
 $\ln\lambda>3.0$&
 10 &
$0.22\pm0.08$&
 $5.2\pm0.3$ \tabularnewline
 $\ln\lambda>3.5$&
 7 &
$0.12\pm0.06$&
 $3.3\pm0.2$ \tabularnewline
\hline
\end{tabular}\end{center}

\caption{Expected number of $\nu_{e}$CC background and signal events in the
$\tau\rightarrow e$ analysis. A multi-dimensional likelihood function
is used as the unique discriminant. Numbers are normalized to 5 years
running of CNGS. Errors in the number of expected events are of 
statistical nature.\label{tab:lkl}}
\end{table}

\subsubsection{The Fisher Discriminant Method}

The Fisher discriminant method \cite{Cowan} is a standard statistical
procedure that, starting from a large number of input variables, allows
us to obtain a single variable that will efficiently distinguish among
different hypotheses. As in the likelihood method, the Fisher discriminant
will contain all the discrimination information. 

The Fisher approach tries to find a linear combination of the following kind

\[
t(\{ x_{j}\})=a_{0}+\sum_{i=1}^{n}a_{i}x_{i}\]
of an initial set of variables $\{ x_{j}\}$ which maximizes

\begin{equation}
J(\{ a_{j}\})=\frac{(\bar{t}_{sig}-\bar{t}_{bkg})^{2}}{(\sigma_{sig}^{2}-\sigma_{bkg}^{2})}\label{eq:fisher}\end{equation}
where $\bar{t}$ is the mean of the $t$ variable and $\sigma$ its
variance. This last expression is nothing but a measure, for the variable
$t$, of how well separated signal and background are. Thus, by maximizing
(\ref{eq:fisher}) we find the optimal linear combination of initial
variables that best discriminates signal from background. The parameters
$a_{j}$ which maximize (\ref{eq:fisher}) can be obtained analytically
by (see \cite{Cowan})

\begin{equation}
a_{i}=W_{ij}^{-1}(\mu_{j}^{sig}-\mu_{j}^{bkg})\label{eq:fisher_coef}\end{equation}
where $\mu_{j}^{sig}$ and $\mu_{j}^{bkg}$ are the mean
in the variable $x_{j}$ for signal and background respectively, and
$W=V_{sig}+V_{bkg}$, being $V$ the covariance matrices.

\paragraph{A Fisher Function for $\nu_\tau$ Appearance Search}

\begin{figure}
\begin{center}\includegraphics[%
  width=10cm,
  keepaspectratio]{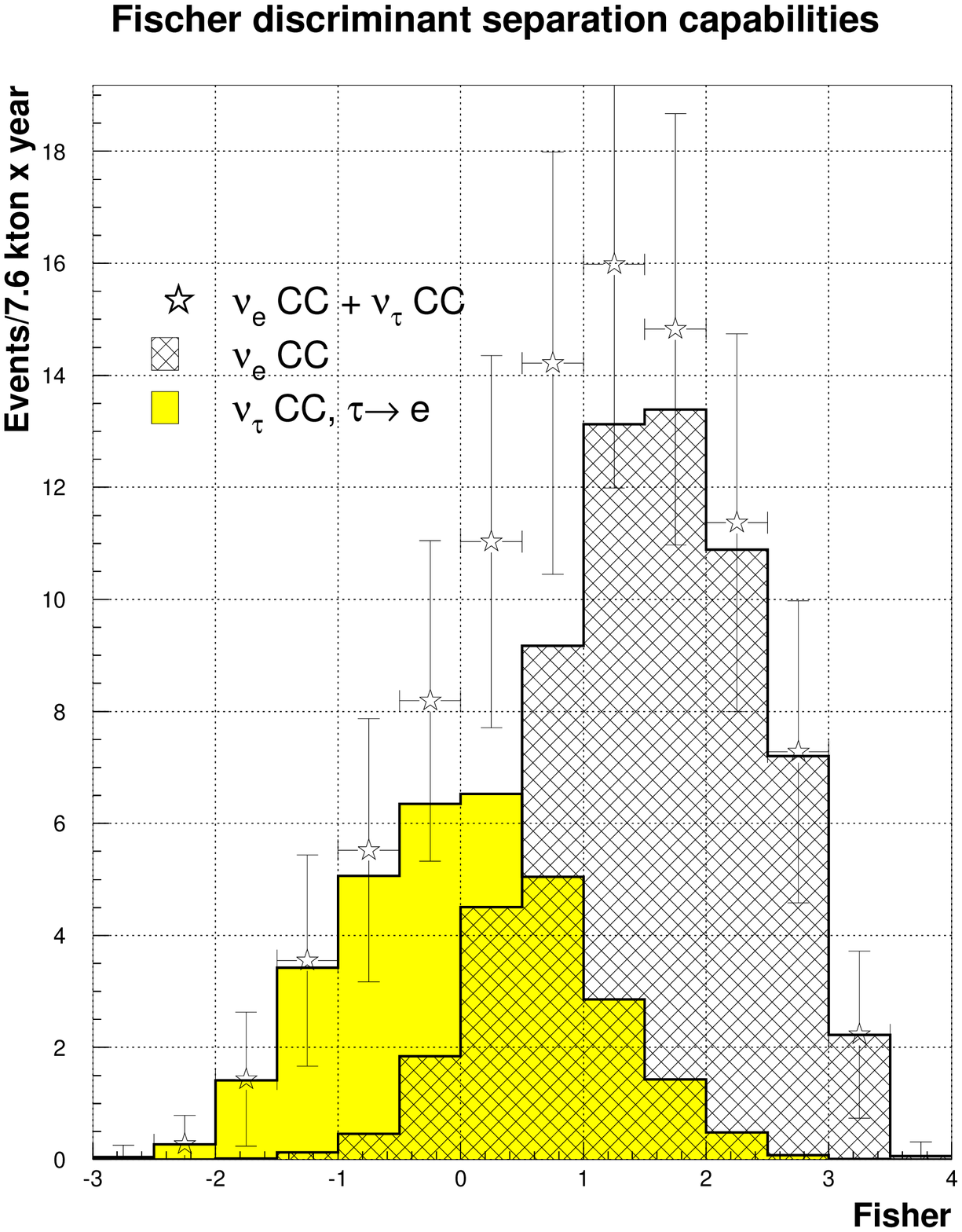}\end{center}

\caption{The Fischer discriminant variable. Error bars in $\nu_{e}$CC + $\nu_{\tau}$CC
sample represent statistical fluctuations in the expected profile
measurements after 5 years of data taking with shared running CNGS
and a 3 kton detector configuration\label{fig:Fischer-discriminant-variable.}}
\end{figure}

\paragraph{}
From the distributions of kinematic variables for $\nu_{\tau}$CC
and $\nu_{e}$CC, we can immediately construct a Fisher function
for a given set of variables. Initially we select the same set of
variables we used for the likelihood approach, namely: $E_{visible}$,
$P_{T}^{miss}$ and $\rho_{l}$. We need only the vector of means
and covariance matrices in order to calculate the optimum Fisher variable 
(equation \ref{eq:fisher_coef}). Distributions are shown in figure
\ref{fig:Fischer-discriminant-variable.}, where the usual normalization
has been assumed. In table \ref{tab:Fisher-tab} values for the expected
number of signal and background events are shown as a function of
the cut on the Fisher discriminant. Since linear correlations among variables
are taken into account, the Fisher discriminant method offers similar results 
to the one obtained using a multi-dimensional likelihood.

\begin{table}
\begin{center}\begin{tabular}{|l|c|c|c|}
\hline 
 &
 $\nu_{\tau}$CC ($\tau\rightarrow e$) &
&
 $\nu_{\tau}$ CC ($\tau\rightarrow e$)\tabularnewline
~~~~~~~~Cuts&
Efficiency&
 $\nu_{e}$CC&
 $\Delta m^{2}=$\tabularnewline
&
(\%)&
&
 $3\times10^{-3}$ eV$^{2}$\tabularnewline
\hline
Initial &
 100 &
 252 &
 50 \tabularnewline
\hline
Fiducial volume &
 65 &
 164 &
 33 \tabularnewline
\hline
 Fisher$>0.5$&
 46 &
$6.9\pm0.3$&
$23.1\pm0.4$\tabularnewline
 Fisher$>0.0$&
 33 &
 $2.4\pm0.2$ &
$16.6\pm0.4$\tabularnewline
\textbf{Fisher}$\mathbf{>-0.27}$&
\textbf{25}&
$\mathbf{1.15\pm0.13}$&
$\mathbf{12.9\pm0.3}$\tabularnewline
 Fisher$>-0.5$&
 20&
$0.60\pm0.10$&
$10.2\pm0.3$\tabularnewline
 Fisher$>-1.0$&
 10 &
 $0.14\pm0.05$ &
 $5.2\pm0.2$ \tabularnewline
\hline
\end{tabular}\end{center}

\caption{Expected number of $\nu_{e}$CC background and signal events in the
$\tau\rightarrow e$ analysis. A Fisher variable is used as the unique
discriminant. Numbers are normalized to 5 years running of CNGS. Errors
in the number of expected events are of statistical nature.\label{tab:Fisher-tab}}
\end{table}

Contrary to what happens with a multi-dimensional likelihood (where the increase in 
the number of discriminating variables demands more Monte-Carlo data and therefore it 
is an extreme CPU-consuming process), the application of the Fisher method to a 
larger number of kinematic variables is straightforward, since the main characteristic 
of the Fisher method is that the final discriminant
can be obtained algebraically from the initial distributions of kinematic
variables. For instance, a Fisher discriminant built out of 9 kinematic variables
($E_{vis}$, $P_{T}^{miss}$, $\rho_{l}$, $P_{T}^{lep}$, $E_{lep}$,
$\rho_{m}$, $Q_{T}$, $m_{T}$, $Q_{lep}$)\footnote{see~\cite{Astier:1999vc} 
for a detailed explanation of the 
variables} predicts for $12.9\pm0.3$
taus a background of $1.17\pm0.14$ $\nu_{e}$ CC events. We conclude
that, for the Fisher method, {\bf increasing the number of variables does not improve the discrimination
power we got with the set $E_{vis}$, $P_{T}^{miss}$, $\rho_{l}$ and therefore these three variables 
are enough to perform an efficient $\tau$ appearance search}.

\subsection{Oscillation Search Using Neural Networks}

In the context of signal vs background discrimination, neural networks
arise as one of the most powerful tools. The crucial point that makes
these algorithms so good is their ability to adapt themselves to the
data by means of non-linear functions.

Artificial Neural Networks have become a promising approach
to many computational applications. It is a mature and well founded
computational technique able to \emph{learn} the natural behaviour
of a given data set, in order to give future predictions or take decisions
about the system that data represent (see \cite{Bishop} and \cite{Haykin}
for a complete introduction to neural networks). During last decade,
neural networks have been widely used to solve High Energy Physics
problems (see \cite{CERN_NN} for a introduction to neural networks
techniques and applications to HEP). Multi-layer perceptrons efficiently
recognize signal features from an, a priori, dominant background environment 
(\cite{Ametller}, \cite{Higgs-Chiapetta}).

We have evaluated the performance offered by neural networks 
when looking for $\nu_{\mu}\rightarrow\nu_{\tau}$
oscillations. As in the case of a multi-dimensional likelihood, a single
valued function will be the unique discriminant. This is obtained
adjusting the free parameters of our neural network model by means
of a \emph{training period}. During this process, the neural network
is taught to distinguish signal from background using a \emph{learning}
data sample.

Two different neural networks models have been studied: the \emph{multi-layer
perceptron} and the \emph{learning vector quantization} self-organized
network. In the following, the results obtained
with both methods are discussed.

\subsubsection{The Multi-layer Perceptron\label{sub:The-Multi-layer-Perceptron}}

\begin{figure}
\begin{center}\includegraphics[%
  width=10cm,
  keepaspectratio]{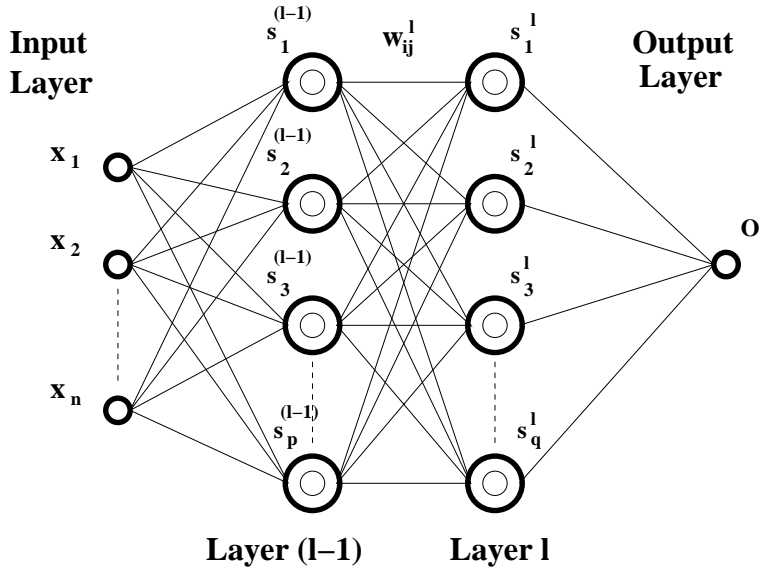}\end{center}

\caption{A general multi-Layer perceptron diagram. The optimal non-linear
function of input variables ($x_{i}$) is constructed using a set
of basic units called \emph{neurons}. Each neuron has two free parameters
that must be adjusted minimizing an error function.\label{fig:mlp}}
\end{figure}

The multi-layer perceptron (MLP) function has a topology based on
different layers of neurons which connect input variables (the variables
that define the problem, also called \emph{feature} variables) with
the output unit (see figure \ref{fig:mlp}). The value (or ''state'')
a neuron has, is a non-linear function of a weighted sum over the
values of all neurons in the previous layer plus a constant, called
bias:

\begin{equation}
s_{i}^{l}=g\,(\sum_{j}\omega_{ij}^{l}s_{j}^{l-1}+b_{i}^{l})\label{eq:mlp_neurons_output}\end{equation}
where $s_{i}^{l}$ is the value of the neuron $i$ in layer $l$;
$\omega_{ij}^{l}$ is the weight associated to the link between neuron
$i$ in layer $l$ and neuron $j$ in the previous layer $(l-1)$;
$b_{i}^{l}$ is a bias defined in each neuron and $g(x)$ is called the 
\emph{transfer function}. The transfer function is
used to regularize the neuron's output to a bounded value between
0 and 1 (or -1,1). 


%
%
%
%

%

In a multi-layer perceptron, a non-linear function is used to obtain
the discriminating variable. Therefore complex correlations among variables 
are taken into account, thus enhancing background rejection capabilities. 

The construction of a MLP implies that several choices must be made
a priori: amount of input variables, hidden layers, neurons per layer, 
number of epochs, etc. The size of the simulated
data set is also crucial in order to optimize the training algorithm performance.
If the training sample is small, it is likely for the MLP to adjust
itself extremely well to this particular data set, thus losing generalization
power (when this occurs the MLP is \emph{over-learning} the data).



\paragraph{Multi-layer Perceptron for $\nu_\tau$ Appearance Search }
\paragraph{} 

As already mentioned in \ref{sub:likelihood}, we fully define our tagging
problem using five variables (three
in the transverse plane and two in longitudinal direction), since they utterly
describe the event kinematics, provided that we ignore the jet structure. 
Initially we build a MLP that contains only three input variables,
and in a latter step we incorporate more variables to see how the
discrimination power is affected. The three chosen variables are $E_{visible}$,
$P_{T}^{miss}$ and $\rho_{l}$. Our election is similar to the one
used for the multi-dimensional likelihood approach. This allows us to
make a direct comparison of the sensitivities provided by the two
methods.

\begin{figure}
\begin{center}\includegraphics[%
  width=10cm,
  keepaspectratio]{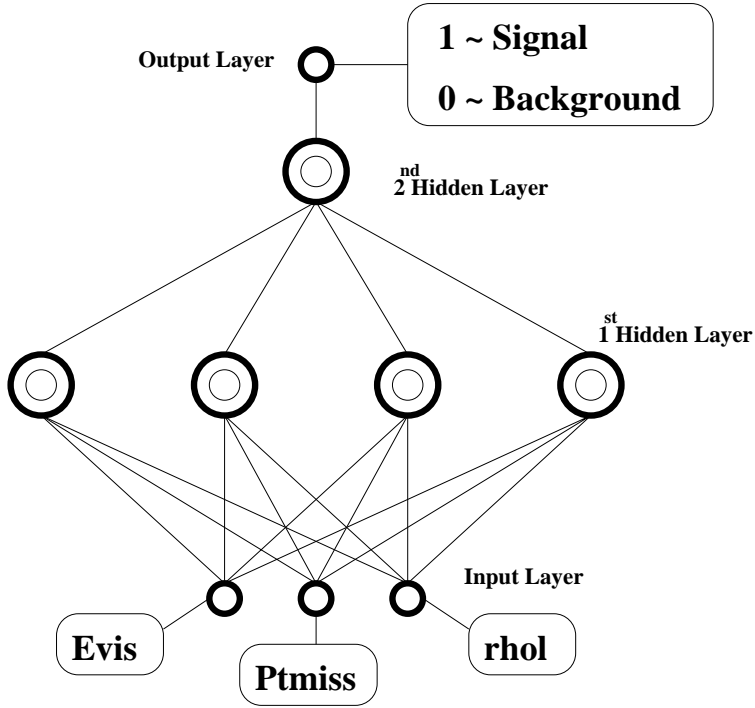}\end{center}

\caption{Chosen topology for the MLP. We feed a two layered
MLP (4 neurons in first layer and 1 in second) with input variables:
$E_{visible}$, $P_{T}^{miss}$ and $\rho_{l}$. \label{fig:mlp_setup}}
\end{figure}


The implementation of the multilayer perceptron was done by means of 
the MLPfit package \cite{PAW mlp}, interfaced in PAW. 
Among the set of different neural network topologies that we studied,  
we saw that the optimal one is made of two hidden
layers with four neurons in the first hidden layer and one in the
second (see figure \ref{fig:mlp_setup}). 

Simulated data was divided in three, statistically independent, subsets
of 5000 events each (consisting of 2500 signal events and an identical amount of background).

The MLP was trained with a first ``learning'' data sample. Likewise,
the second ``test'' data set was used as a training sample to check
that over-learning does not occur. Once the MLP is set, the evaluation
of final efficiencies is done using the third independent data sample
(namely, a factor 40 (75) larger than what is expected for background
(signal) after five years of CNGS running with a 3 kton detector).

\begin{figure}
\begin{center}\includegraphics[%
  width=10cm,
  keepaspectratio]{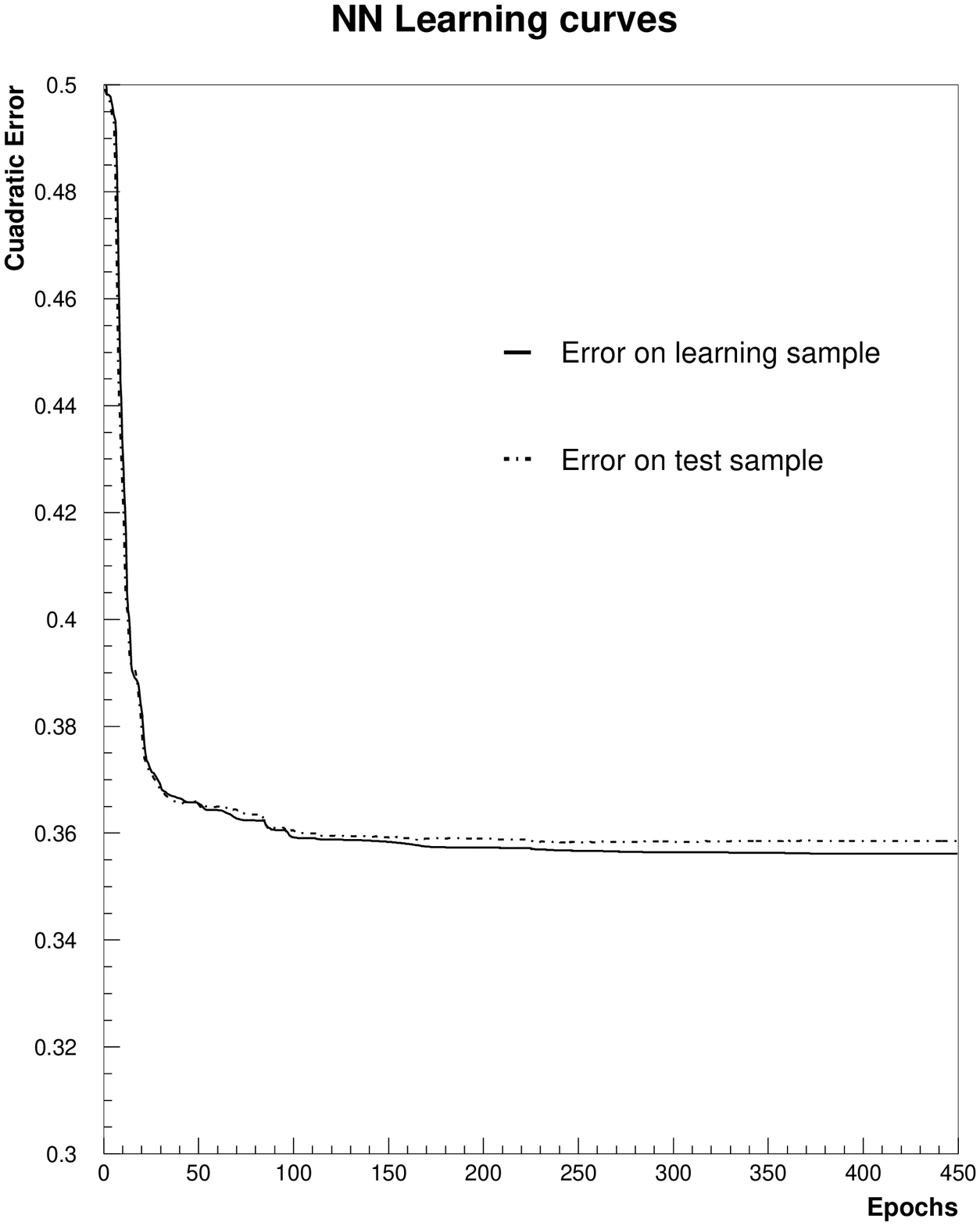}\end{center}

\caption{Learning curves for the MLP. The neural network is trained for 450
epochs in order to reach a stable minimum. The solid line represents
the error on training sample, the dashed line is the error on the
test sample. Both lines run almost parallel: no over-learning occurs.\label{fig:learning_curves}}
\end{figure}

\begin{figure}
\begin{center}\includegraphics[%
  width=12cm,
  keepaspectratio]{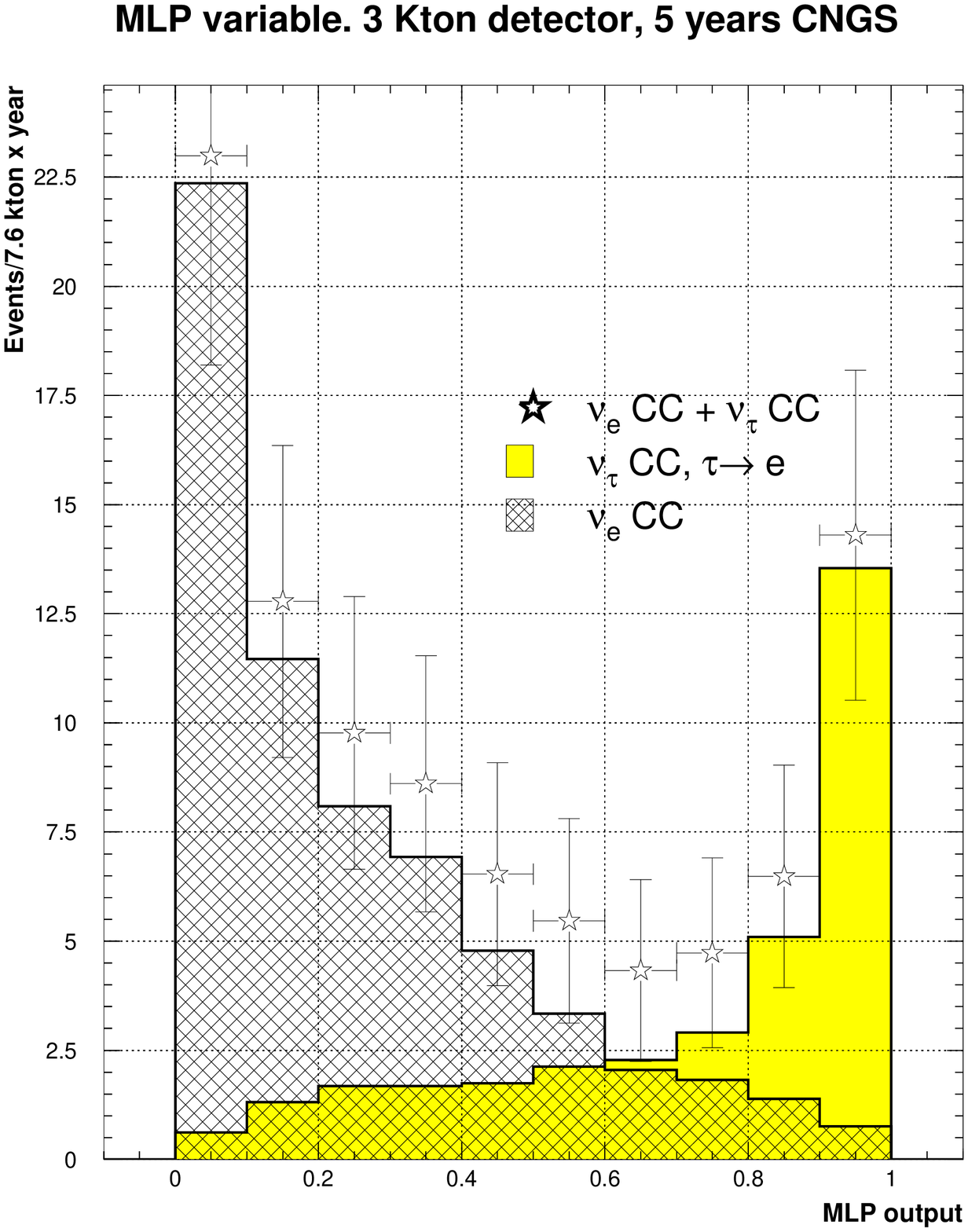}\end{center}

\caption{Multi-layer perceptron output for $\nu_{\tau}$CC ($\tau\rightarrow e$)
and $\nu_{e}$CC events. We see how signal events accumulate around
1 while background peaks at 0.\label{fig:MLP-output.} Only statistical errors are plotted.}
\end{figure}

Error curves during learning are shown in
figure \ref{fig:learning_curves} for training and test samples. We
see that even after 450 epochs, over-learning does not take place.
Final distributions in the multi-layer perceptron discriminating variable
can be seen in figure \ref{fig:MLP-output.}.

\begin{figure}
\begin{center}\includegraphics[%
  width=12cm,
  keepaspectratio]{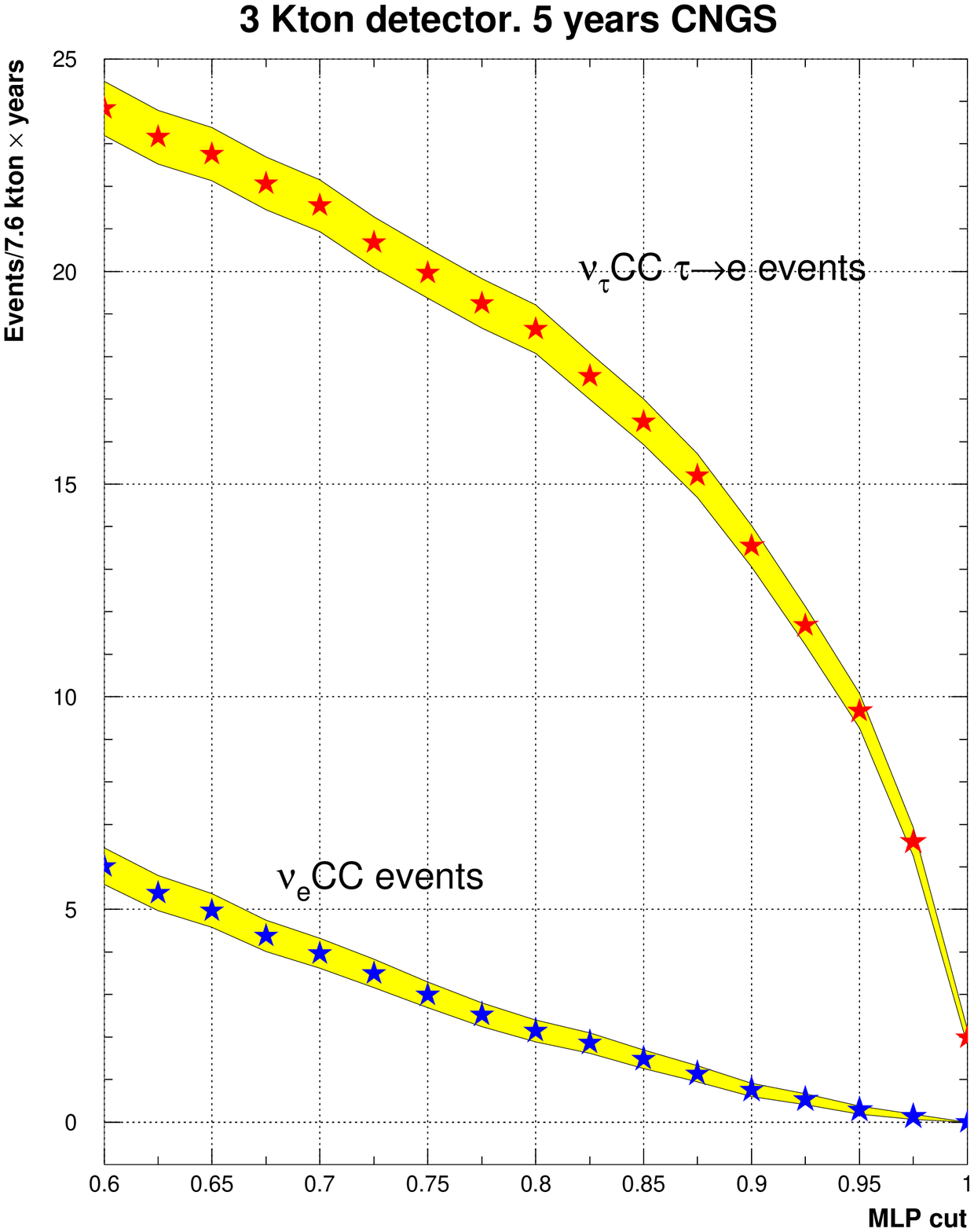}\end{center}

\caption{Number of signal and background events after 5 years of running CNGS
as a function of the MLP cut. Shadowed zones correspond to statistical
errors.\label{fig:mlp_sig_bac_cut}}
\end{figure}

\begin{figure}
\begin{center}\includegraphics[%
  width=12cm,
  keepaspectratio]{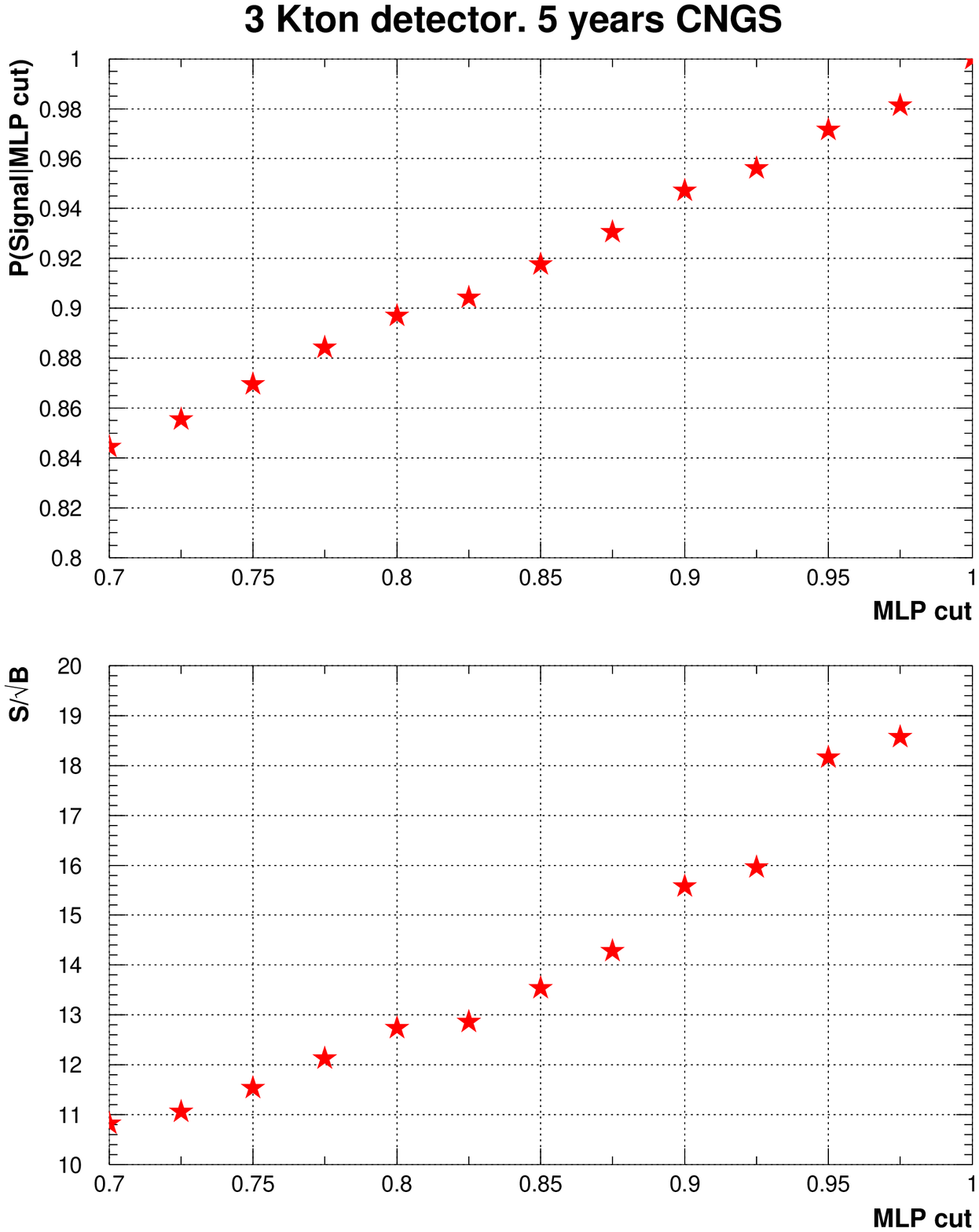}\end{center}

\caption{(Top) Probability of an event belonging to a region in input variable
space characterized by MLP $>cut$ of being a signal event. (Bottom)
Statistical significance of signal events as a function of the cut
in MLP.\label{fig:mlp_eff}}
\end{figure}

\begin{figure}
\begin{center}\includegraphics[%
  width=12cm,
  keepaspectratio]{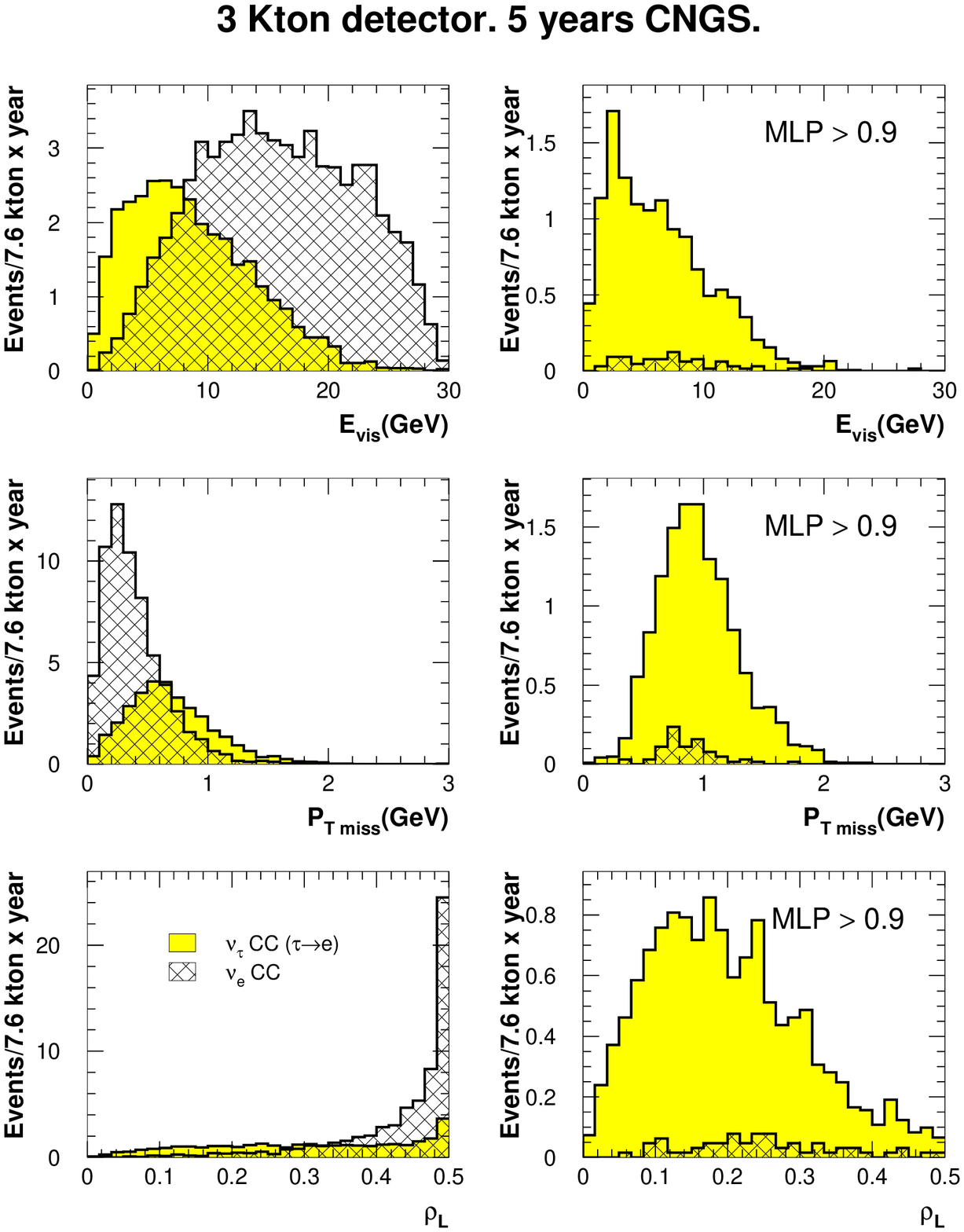}\end{center}

\caption{Kinematic variables before (left histograms) and after (right histograms)
cuts are applied based on the MLP output. We see how the MLP has learnt
that signal events favour low $E_{visible}$, high $P_{T}^{miss}$
and low $\rho_{l}$ values.\label{fig:mlp_cutted_vars}}
\end{figure}

Figure \ref{fig:mlp_sig_bac_cut} shows the number of signal and background
expected after 5 years of data taking as a function of the cut in
the MLP variable. In figure \ref{fig:mlp_eff} we represent the probability
of an event, falling in a region of the input space characterized
by MLP output $>$ cut, to be a signal event (top plot), and the
statistical significance as a function of the MLP cut (bottom plot).
Background rejection has been optimized since a cut based on the MLP output
variable can select regions of complicated topology in the kinematic
hyperspace, given that now complex correlations are taken into account (see figure \ref{fig:mlp_cutted_vars}).

  Selecting MLP $>0.91$ (overall $\tau$ selection efficiency
$=25$\%), the probability that an event falling in this region is
signal amounts to $\sim0.95$. For 5 years of running CNGS and a 3 kton
detector, we expect a total amount of $12.9\pm0.5$ $\nu_{\tau}$CC
($\tau\rightarrow e$) events and $0.66\pm0.14$ $\nu_{e}$CC events.
Table \ref{tab:mlp} summarizes as a function of the applied MLP cut
the expected number of signal and background events.

  If we compare the outcome of this approach with the one
obtained in section \ref{sub:likelihood}, we see that for the same
$\tau$ selection efficiency, the multi-dimensional likelihood expects
$1.1\pm0.2$ background events. Therefore, for this particular cut,
the MLP achieves a 60\% reduction in the number of expected $\nu_{e}$
CC events.

\begin{table}
\begin{center}\begin{tabular}{|l|c|c|c|}
\hline 
 &
 $\nu_{\tau}$CC ($\tau\rightarrow e$) &
&
 $\nu_{\tau}$ CC ($\tau\rightarrow e$) \tabularnewline
~~~~~~~~Cuts&
Efficiency&
$\nu_{e}$CC&
 $\Delta m^{2}=$\tabularnewline
&
(\%) &
&
 $3\times10^{-3}$ eV$^{2}$\tabularnewline
\hline
Initial &
 100 &
 252 &
 50 \tabularnewline
\hline
Fiducial volume &
 65 &
 164 &
 33 \tabularnewline
\hline
MLP $>0.70$&
 42 &
 $4.0\pm0.4$ &
 $21.4\pm0.6$\tabularnewline
MLP $>0.75$&
 40 &
$3.0\pm0.3$&
 $19.9\pm0.6$\tabularnewline
MLP $>0.80$&
 37 &
$2.1\pm0.3$&
$18.6\pm0.5$\tabularnewline
MLP $>0.85$&
 33 &
 $1.5\pm0.2$ &
$16.4\pm0.5$\tabularnewline
\textbf{MLP $>0.90$}&
 27&
$0.76\pm0.15$&
$13.5\pm0.5$\tabularnewline
{\bf MLP} $\mathbf{>0.91}$&
\textbf{25}&
$\mathbf{0.66\pm0.14}$&
$\mathbf{12.9\pm0.5}$\tabularnewline
MLP $>0.95$&
 19 &
 $0.28\pm0.09$ &
 $9.6\pm0.4$ \tabularnewline
MLP $>0.98$&
12&
$0.09\pm0.05$&
$5.8\pm0.3$\tabularnewline
\hline
\end{tabular}\end{center}

\caption{Expected number of background and signal events when a multi-layer
perceptron function is used as the unique discriminant. Numbers are
normalized to 5 years running of CNGS. Errors in the number of events
expected are of statistical nature.\label{tab:mlp}}
\end{table}

As we did for the Fisher method, we studied if the sensitivity given by the MLP increases 
when a larger number of input variables is used.  
Even though the number of complex correlations among variables is larger, the change in the final
sensitivity is negligible. Once again, all the discrimination power is provided by
$E_{visible}$, $P_{T}^{miss}$ and $\rho_{l}$. The surviving background
can not be further reduced by increasing the dimensionality of the
problem. 

  Since an increase on the number of input variables
does not improve the discrimination power of the multi-layer perceptron,
we tried to enhance signal efficiency following a different approach:
optimizing the set of input variables by finding new linear combinations
of the original ones (or functions of them like squares, cubes, etc). 

  To this purpose, using the fast computation capabilities
of the Fisher method, we can operate in a systematic way in order
to find the most relevant feature variables. Starting from an initial
set of input variables, the algorithm described in \cite{Roe} tries
to gather all the discriminant information in an smaller set of optimized
variables. These last variables are nothing but successive Fisher
functions of different combinations of the original ones.
In order to allow not only linear transformations, we can add non-linear
functions of the kinematic variables like independent elements of
the initial set.

  We performed an analysis similar to the one described in \cite{Roe}, using
5 initial kinematic variables ($E_{vis}$, $P_{T}^{miss}$, $\rho_{l}$,
$P_{T}^{lep}$ and $E_{lep}$) plus their cubes and their exponentials
(in total 15 initial variables). At the end, we chose a smaller subset
of six optimized Fisher functions that we use like input features
variables for a new multi-layer perceptron. 

  The MLP analysis with six Fisher variables does not enhance
the oscillation search sensitivity that we got with the three usual
variables $E_{vis}$, $P_{T}^{miss}$and $\rho_{l}$. We therefore
conclude that {\bf neither the increase on the number of features variables
nor the use of optimized linear combinations of kinematic variables
as input, enhances the sensitivity provided by the MLP}. 

The application of statistical techniques able to find complex
correlations among the input variables is the only way to enhance
background rejection capabilities. In this respect, neural networks
are an optimal approach.

\subsubsection{  Self Organizing Neural Networks: LVQ Network}

\begin{figure}
\begin{center}\includegraphics[%
  width=10cm,
  keepaspectratio]{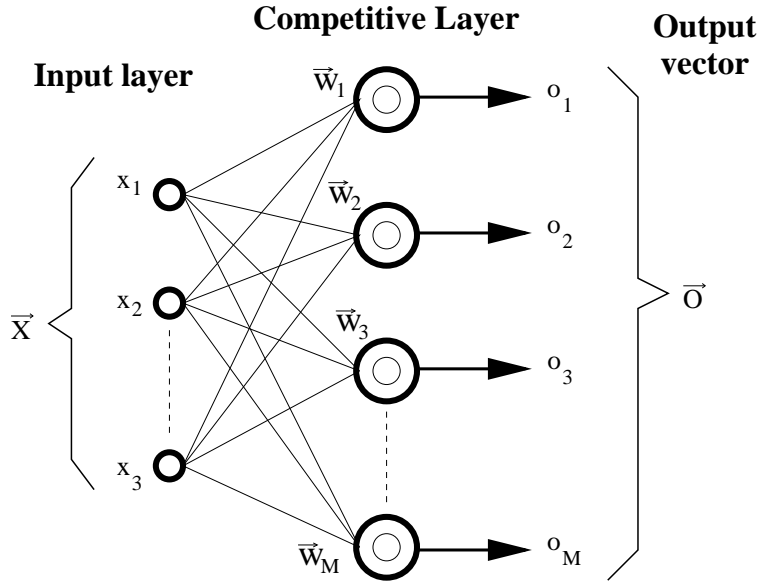}\end{center}

\caption{Schematic diagram of the general topology for a \emph{self-organized}
neural network. Neurons in the competitive layer are connected with
each one of the input nodes. 
\label{fig:Self-organized-network}}
\end{figure}

  A \emph{self-organizing} (SO) network operates in a different
way than a multi-layer perceptron does. These networks have the ability
to organize themselves according to the ``natural structure'' of
the data. They can learn to detect regularities and correlations in
their input and adapt their future response to that input accordingly.
A SO network usually has, besides the input, only one layer of neurons
that is called \emph{competitive layer} (see figure \ref{fig:Self-organized-network}). 
Neurons in the competitive layer are able to learn the structure
of the data following a simple scheme called \emph{competitive self-organization}
(see \cite{CERN_NN}), which ``moves'' the basic units (neurons)
in the competitive layer in such a way that they imitate the natural
structure of the data.

Competitive self-organization is an unsupervised
learning algorithm, however for classification purposes one can improve the
algorithm with supervised learning in order to fine tune final positions
of the neurons in the competitive layer. This is called \emph{learning
vector quantization} (LVQ) (for further details refer to \cite{Bishop, CERN_NN}). 
An important difference with respect to
the multi-layer perceptron approach is that in LVQ we always get a
discrete classification, namely, an event is always classified in
one of the classes. The only thing one can estimate is the degree
of belief in the LVQ choice.

\paragraph{LVQ Network for $\nu_\tau$ Appearance Search }
\paragraph{} 

\begin{figure}
\begin{center}\includegraphics[%
  width=12cm,
  keepaspectratio]{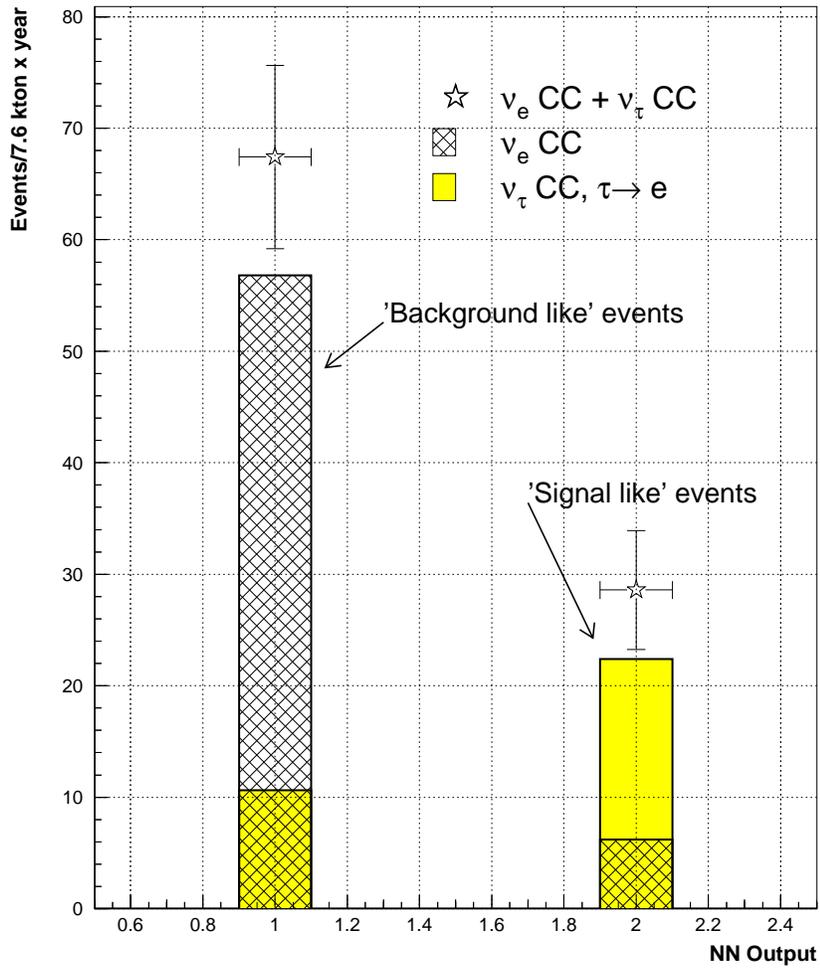}\end{center}

\caption{LVQ neural network separation capabilities. In competitive self-organized
networks a discrete decision is always issued: \emph{signal like}
events are labeled with 2 and \emph{background like} with 1.\label{fig:LVQ-neural-network}}
\end{figure}

We use once more $E_{visible}$, $P_{T}^{miss}$ and $\rho_{l}$
as discriminating variables inside the input layer. A LVQ network with 10 neurons has been
trained with samples of 2500 events for both signal and background.
Given that, before any cut, a larger background sample is expected,
we have chosen an asymmetric configuration for the competitive layer.
Out of 10 neurons, 6 were assigned to recognize background events,
and the rest were associated to the signal class. After the neurons
are placed by the training procedure, the LVQ network is fed with
a larger and statistically independent data sample consisting of 6000 signal and
background events. The output provided by the network is plotted in
figure \ref{fig:LVQ-neural-network}. We see how events are classified
in two independent classes: \emph{signal like} events (labeled with
2) and \emph{background like} ones (labeled with 1). 68\% of $\nu_{\tau}$CC
($\tau\rightarrow e$) events and 10\% of $\nu_{e}$CC events, occurring
in fiducial volume, are classified as \emph{signal like} events.
This means a $\tau$ efficiency around 45\% with respect to the tau
events generated in active LAr. For the same $\tau$ efficiency, the
multi-layer perceptron only misclassified around 8\% of $\nu_{e}$CC
events.

  Several additional tests have been performed with LVQ networks,
by increasing the number of input variables and/or the number of neurons
in the competitive layer. However, 
we observed no improvement on the separation capabilities. 
For instance, a topology with 16 feature neurons in
the competitive layer and 4 input variables (we add the transverse
lepton momentum) leads to exactly the same result. 

  The simple geometrical interpretation of this kind of neural
networks supports our statement that the addition of new variables
to the original set \{$E_{visible}$, $P_{T}^{miss}$ , $\rho_{l}$\}
does not enhance the discrimination power: the bulk of signal and
background events are not better separated when we increment the dimensionality
of the input space.

\paragraph{  Combining MLP with LVQ}
\paragraph{}

  We have seen that LVQ networks returns a discrete output.
The whole event sample is classified by the LVQ in two classes: \emph{signal-like}
and \emph{background-like.} We can use the classification of a LVQ
as a \emph{pre-}classification for the MLP. A priori, it seems reasonable
to expect an increase on the oscillation search sensitivity if we
combine the LVQ and MLP approaches. The aim is to evaluate how much
additional background rejection, from the \emph{contamination inside
the signal-like} sample, can be obtained by means of a MLP. 

\begin{figure}
\begin{center}\includegraphics[%
  width=12cm,
  keepaspectratio]{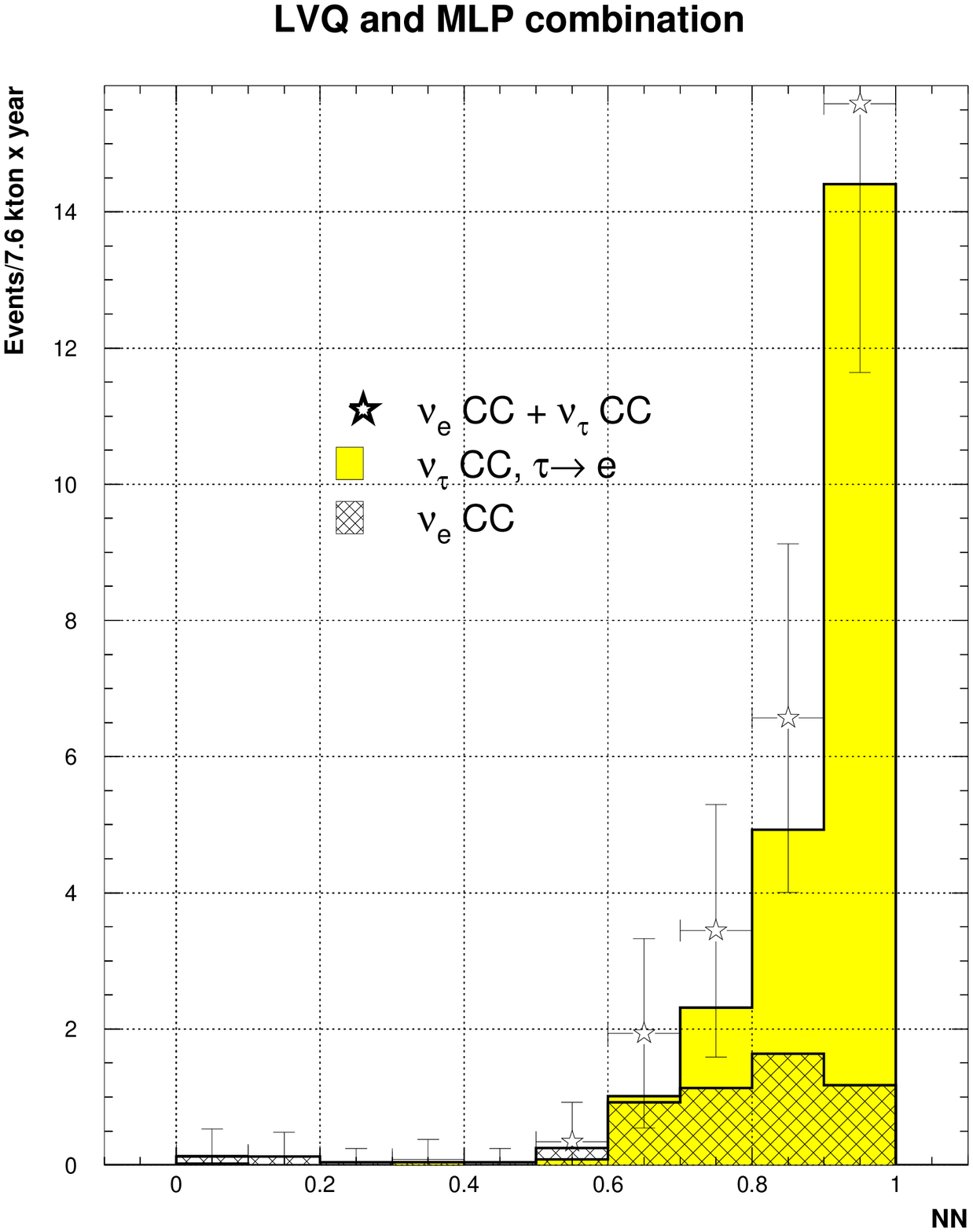}\end{center}

\caption{LVQ and MLP networks combined. Distributions are given in the continuous
MLP variable. Only events labeled by LVQ with 2 (\emph{signal like})
have been used for the analysis.\label{fig:LVQ-and-MLP}}
\end{figure}

  We present in figure \ref{fig:LVQ-and-MLP} the MLP output
for events classified as \emph{signal-like} by the LVQ network (see
figure \ref{fig:LVQ-neural-network}). Applying a cut on the MLP output
such that we get $12.9$ signal events (our usual reference
point of 25\% $\tau$ selection efficiency), we get $0.82\pm0.19$
background events, similar to what was obtained with the MLP approach
alone. This outcome conclusively shows that, contrary to our a priori
expectations, an event pre-classification, by means of a learning
vector quantization neural network, does not help improving the discrimination
capabilities of a multi-layer perceptron.

\section{$\nu_\tau$ Discovery Potential}

We have studied several pattern recognition techniques applied to
the particular problem of searching for $\nu_{\mu}\rightarrow\nu_{\tau}$
oscillations. Based on \emph{discovery criteria}, similar 
to the ones proposed in \cite{discovery} 
for statistical studies of prospective nature, we try to quantify how much 
the statistical relevance of the $\tau$ signal varies depending 
on the statistical method used.

We define $\mu_{S}$ and $\mu_{B}$ as the average number of expected signal
and background events, respectively. With this notation, we impose two 
conditions to consider that a signal is statistically significant:

\begin{enumerate}
\item We require that the probability for a background fluctuation, giving
a number of events equal or larger than $\mu_{S}+\mu_{B}$,
be smaller than $\epsilon$ (where $\epsilon$ is $5.733\times10^{-7}$, 
the usual $5\sigma$ criteria applied for Gaussian distributions).
\item We also set at which confidence level $(1-\delta)$, the distribution of the total
number of events with mean value $\mu_{S}+\mu_{B}$ fulfills the background 
fluctuation criteria stated above.
\end{enumerate}

For instance, if $\delta$ is $0.10$ and $\epsilon$ is $5.733\times10^{-7}$,
we are imposing that $90\%$ of the times we repeat this experiment, we will
observe a number of events which is $5\sigma$ or more above the background expectation.

For all the statistical techniques used, we fix $\delta$=$0.10$
and $\epsilon$=$5.733\times10^{-7}$. In this way we can compute the minimum
number of events needed to establish that, in our particular example, 
a direct $\nu_{\mu}\rightarrow\nu_{\tau}$ oscillation has been observed.

In table \ref{tab:discovery} we compare the number of signal and
background events obtained for the multi-layer perceptron and the multi-dimensional
likelihood approaches after 5 years of data taking with a 3 kton detector.
We also compare the minimum exposure needed in order to have a statistically 
significant signal.
The minimum exposure is expressed in terms of a scale factor $\alpha$, 
where $\alpha=1$ means a total exposure of $11.75$~kton$\times$year. 
For the multi-layer perceptron approach ($\alpha=0.86$), a statistically 
significant signal can be obtained 
after a bit more than four years of data taking. 
On the other hand, the multi-dimensional likelihood approach requires 
5 full years of data taking. Therefore, when applied to 
the physics quest for neutrino oscillations, neural 
network techniques are more performant than classic statistical methods.
\begin{table}
\begin{center}\begin{tabular}{|l|c|c|}
\cline{2-2} \cline{3-3} 
\multicolumn{1}{c|}{}&
\textbf{Multi-layer}&
\textbf{Multi-dimensional}\tabularnewline
\multicolumn{1}{c|}{}&
\textbf{Perceptron}&
\textbf{Likelihood}\tabularnewline
\hline
\multicolumn{1}{|l|}{\textbf{\# Signal}}&
12.9&
12.9\tabularnewline
\hline 
\textbf{\# Background}&
0.66&
1.1\tabularnewline
\hline 
\textbf{$\alpha$~factor}&
\textbf{0.86}&
\textbf{1.01}\tabularnewline
\hline
\end{tabular}\end{center}

\caption{Number of signal and background events for
the multi-layer perceptron and the multi-dimensional likelihood approaches.
Numbers are normalized to 5 years of data taking in shared CNGS running mode 
and a 3 kton detector configuration. 
The last row displays the scale factor $\alpha$ needed to compute the minimum 
exposure fulfilling the discovery criteria described in the text.\label{tab:discovery}}
\end{table}

\section{Conclusions}

  We have considered the general problem of $\nu_\mu\to\nu_{\tau}$ 
oscillation search 
based on kinematic criteria to assess the performance 
of several statistical pattern recognition methods.

Two are the main conclusions of this study: 
\begin{itemize}
\item An optimal discrimination power 
is obtained using only the following variables:
$E_{visible}$, $P_{T}^{miss}$ and $\rho_{l}$ and their correlations.
Increasing the number of variables (or combinations of variables)
only increases the complexity of the problem, but does not result
in a sensible change of the expected sensitivity. 
\item Among the set of statistical methods considered, the multi-layer
perceptron offers the best performance. 
\end{itemize}

\begin{figure}[H]
\begin{center}\includegraphics[%
  width=10.5cm,
  keepaspectratio]{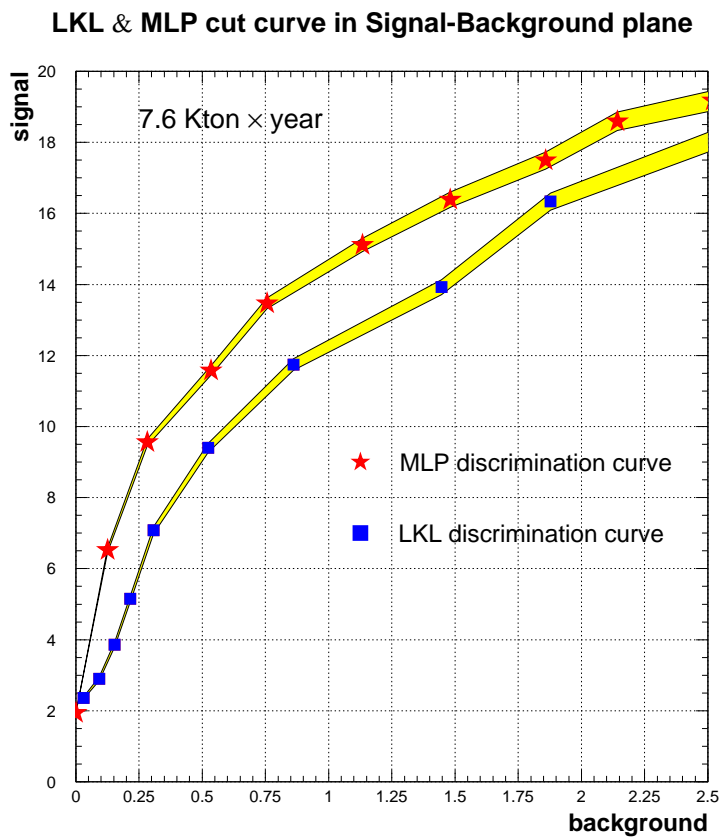}\end{center}

\caption{Multi-layer perceptron vs multi-dimensional likelihood. We assume a 7.6 
Kton$\times$year
exposure. The shadowed area shows the statistical error.\label{fig:mlpvslkl}}
\end{figure}

As an example, we have considered the case of the CNGS beam and 
$\nu_\tau$ appearance search (for the $\tau\to e$ decay channel) using a 
very massive (3 kton) Liquid Argon TPC detector. Figure \ref{fig:mlpvslkl}
compares the discrimination capabilities of multi-dimensional likelihood
and multi-layer perceptron approaches. We see that, for the low background
region, the multi-layer perceptron gives the best sensitivity. For
instance, choosing a $\tau$ selection efficiency of 25\% as a reference
value, we expect a total of $12.9\pm0.5$ $\nu_{\tau}$CC ($\tau\rightarrow e$)
signal and $0.66\pm0.14$ $\nu_{e}$CC background. Compared to multi-dimensional
likelihood predictions, this means a 60\% reduction on the number
of expected background events. Hence, using a multi-layer perceptron, 
fours years of data taking will suffice to get a statistically significant 
signal, while five years are needed when the search approach is based on 
a multi-dimensional likelihood.

\section*{Acknowledgments}

We thank Paola Sala for discussions and help with Monte-Carlo generation.

\end{document}